\documentclass[%
 reprint,
 amsmath,amssymb,
 aps,
]{revtex4-1}

\usepackage{graphicx}% Include figure files
\usepackage{dcolumn}% Align table columns on decimal point
\usepackage{bm}% bold math
\begin{document}

\preprint{APS/123-QED}

\title{Saturation and alternate pathways in four-wave mixing in rubidium}% Force line breaks with \\

\author{E. Brekke}
 \email{erik.brekke@snc.edu}
\author{N. Swan}%

\affiliation{%
 St. Norbert College, Department of Physics,
 De Pere, WI 54115
}%

\date{\today}% It is always \today, today,

\begin{abstract}
We have examined the frequency spectrum of the blue light generated via four-wave mixing in a rubidium vapor cell inside a ring cavity.  At high atomic density and input laser power, two distinct frequency components separated by $116 \pm 4$ MHz are observed, indicating alternate four-wave mixing channels through the $6p_{3/2}$ hyperfine states.  The dependence of the generated light on excitation intensity and atomic density are explored, and indicate the primary process has saturated.  This saturation results when the excitation rate through the 6p state becomes equal to the rate through the 5p state, giving no further gain with atomic density while a quadratic intensity dependence remains.
\end{abstract}

\pacs{Valid PACS appear here}% PACS, the Physics and Astronomy
\maketitle

%\tableofcontents

\section{\label{sec:level1}Introduction}

Nonlinear optical processes can result in a wide range of phenomena in atomic media.  Four-wave mixing in particular has drawn considerable interest in the areas of single photon sources \cite{Ripka:18}, quantum information and slow light \cite{Camacho:09,Radnaev:10}, entanglement and two mode squeezing \cite{Lawrie:16,Marino:09}, Rydberg states \cite{deMelo:14,Brekke:08}, the transfer of orbital angular momentum \cite{Akulshin:16, Offer:18}, and frequency up-conversion.

Frequency up-conversion through parametric four-wave mixing has now been demonstrated in a number of environments, including near resonant cw excitation in rubidium \cite{Zibrov:02,Meijer:06,Akulshin:12b, Vernier:10} and cesium \cite{Schultz:09}, and pulsed or cw excitation far detuned from the intermediate state \cite{Wunderlich:90,Lopez:16,Brekke:13, Kitano:17}.  Extensive work has been done to understand the frequency characteristics \cite{Akulshin:12, Akulshin:14,Brekke:15}, observe the infrared emission \cite{Akulshin:14b,Sell:14}, and examine competing pathways and processes \cite{Akulshin:15,Gai:16,Prajapati:18}.

In this paper we examine the parametric four-wave mixing process in the regime where saturation of the primary channel occurs due to interference between alternate excitation pathways.  This results in a limit to the effectiveness of higher atomic densities and a reduced gain with input intensity.  In addition, the saturation of the primary channel leads to four-wave mixing on an alternate pathway through different excited hyperfine states becoming relatively more significant, resulting in two distinct frequency peaks in the blue light separated by $116 \pm4$ MHz in $^{87}$Rb.  The onset of saturation and its dependence on excitation parameters can play a large role in the resulting frequency converted light.

\section{\label{sec:level2}Experimental Method and Setup}

\begin{figure}[htbp]
\centering
\includegraphics[width=7.8 cm]{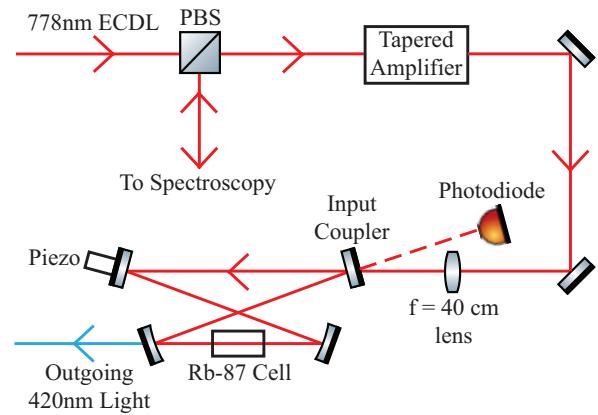}
\caption {A simplified version of the experimental setup.  A single ECDL laser on the $5s_{1/2}\rightarrow5d_{5/2}$ transition is amplified and focused through a $^{87}$Rb cell inside a ring cavity.  The resulting 420 nm light is sent into a 1.5 GHz FSR Fabry-Perot cavity.}
\label{fig:ExpSetup}
\end{figure}

\begin{figure}[htbp]
\centering
\includegraphics[width=8.0cm]{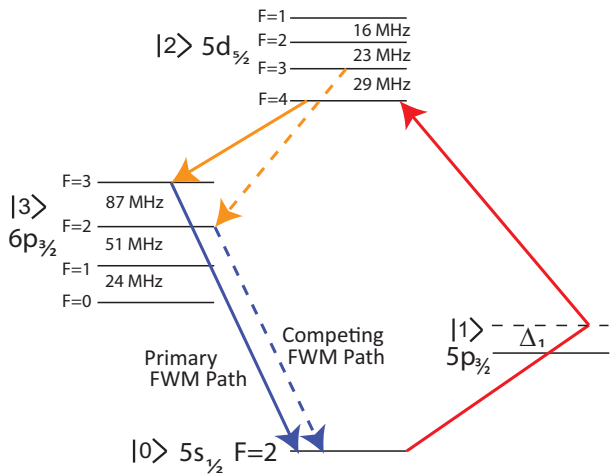}
\caption {Energy levels involved in the four-wave mixing process in $^{87}$Rb. Two-photon excitation is accomplished using a single 778 nm ECDL on the $5s_{1/2}\rightarrow5d_{5/2}$ transition detuned 1 THz from the 5p state.  The resulting process produces coherent and collimated beams at 5.23 $\mu$m and 420 nm. The primary four-wave mixing path goes through the $6p_{3/2}$ F=3 state, with an alternate path through the $6p_{3/2}$ F=2 state.  }
\label{fig:levels}
\end{figure}

Our experimental setup is schematically illustrated in Fig. \ref{fig:ExpSetup}. A single extended cavity diode laser (ECDL) at 778 nm excites the two photon $5s_{1/2}\rightarrow5d_{5/2}$ transition in rubidium.  The frequency control and tapered amplifier system have been described previously \cite{Brekke:13}.  Amplified spontaneous emission and four-wave mixing in rubidium result in generated beams at 5.23 $\mu$m and 420 nm.  The relevant energy levels are shown in Fig. \ref{fig:levels}.

The use of a ring cavity as described in \cite{Brekke:17} allows us to attain blue powers in excess of 100 $\mu$W.  At these power levels saturation, power broadening, and alternate four-wave mixing pathways should play a significant role.  An amplified photodiode measures blue power generated, while a scanning Fabry-Perot cavity with FSR 1.5 GHz is used to examine the frequency characteristics of the resulting blue light.   The beam generated at 5.23 $\mu$m is absorbed by the glass cell, but has been observed elsewhere through the use of Sapphire windows \cite{Akulshin:14b}.

%\begin{figure}[htbp]
%\centering
%\includegraphics[width=8.2cm]{braodening}
%\caption {Still Needed - A graph of the linewidth of the 420 nm as a function of power produced.  Excitation was done at +500 MHz from the $5s_{1/2} F=2\rightarrow5d_{5/2} F=4$ transition to minimize the effect of absorption.}
%\label{fig:broadening}
%\end{figure}

\section{Alternate Transitions in Four-Wave Mixing}

\begin{figure}[htbp]
\centering
\includegraphics[width=8.2cm]{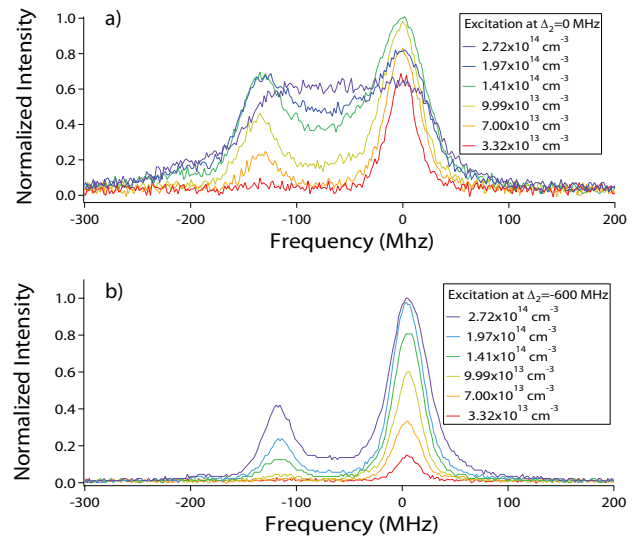}
\caption {Frequency spectrum of the 420 nm light for a variety of atomic densities. a) Excitation on resonance with the $5s_{1/2}$ $F=2\rightarrow5d_{5/2}$ F=4 transition. b) Excitation detuned -600 MHz from the $5s_{1/2}$ $F=2\rightarrow5d_{5/2}$ F=4 transition. The appearance of a second peak at -116 MHz from the primary peak becomes significant at high densities, and continues to grow while the primary peak growth slows. Both data sets were taken with a circulating intensity of $1.2\times10^{10}$ $\frac{W}{m^2}$}    
\label{fig:densityspectrums}
\end{figure}

Four-wave mixing can be done over a wide range of excitation frequencies due to the high temperature of the cell and resulting Doppler width.   The excitation laser is tuned near the $5s_{1/2}$ $F=2\rightarrow5d_{5/2}$ F=4 transition in $^{87}$Rb, with the frequency controlled using two photon spectroscopy.  Excitation detuned from this transition selects a non-zero velocity class of atoms which are Doppler shifted into resonance.  Only one 420 nm frequency mode is observed at low excitation rates and atomic densities, corresponding to four-wave mixing along the $5d_{5/2}$ F=4$\rightarrow6p_{3/2}$ F=3$\rightarrow5s_{1/2}$ F=2 path.  However, as the generated power increases, a second frequency mode is observed in the 420 nm light, as shown in Fig. \ref{fig:densityspectrums}.  

We have examined the dependence of this secondary peak on atomic density, excitation intensity, and detuning.  The spectrum of the 420 nm light in the Fabry-Perot interferometer is shown in Fig. \ref{fig:densityspectrums}a for a variety of atomic densities.  Here excitation occurs on resonance with a circulating intensity of $1.2\times10^{10}$ $\frac{W}{m^2}$.  As density increases, the primary peak ceases to increase in power, while a second peak appears and  continues to grow, changing the relative size of the peaks.  At high densities, the power at the primary peak even decreases, suggesting the importance of absorption inside the cell.

%As has been observed previously\cite{Brekke:17}, the power at 420 nm saturates at high atomic densities.  This can be seen in Fig. \ref{fig:densityspectrumson}, where the primary peak saturates.  A second peak appears and  continues to grow at high densities, changing the relative size of the peaks.  At high densities, the power at the primary peak even decreases, suggesting the importance of absorption inside the cell.

%Do we want a figure showing saturation parameters again?

In order to minimize the effect of absorption, the blue spectra were taken while the excitation process was detuned -600 MHz from the $5s_{1/2}$ $F=2\rightarrow5d_{5/2}$ F=4 transition.   As previously shown \cite{Brekke:15}, detuned excitation results in the detuning of the 420 nm beam.  Figure \ref{fig:densityspectrums}b shows the resulting spectra for a variety of densities, again with circulating intensity of $1.2\times10^{10}$ $\frac{W}{m^2}$.  Here the same general trends are confirmed, with growth of the primary peak slowing relative to the secondary peak, while the effect of absorption is minimized.  

%\color{red} To further investigate something, the blue spectra was measured for a range of different excitation detunings.  It is hard to say what we learned here \color{black}

Fitting the Fabry-Perot peaks allowed us to determine the width of the peaks and their separation.  The separation between the peaks was consistent for a large range of densities and intensities at $116 \pm 4$ MHz.  Our ring cavity has a FSR of 500 MHz, so this spacing is not related to cavity modes.  We hypothesize that the secondary peak is caused by an alternate hyperfine path for the four-wave mixing process \cite{Meijer:06}.  At the high temperatures of the cell, a large number of atoms would have the correct velocity to have the $5d_{5/2}$ F=3 state Doppler shifted into resonance.  This state could then decay to the $6p_{3/2}$ F=2 state, resulting in a different blue frequency during the final step.  The relevant hyperfine levels are shown in Fig. \ref{fig:levels}.  In principle, there could be four-wave mixing paths through the other two $5d_{5/2}$ hyperfine states with different velocity classes, but this was not observed with the densities and powers available. 

In addition to the 87 MHz shift for the 6p hyperfine states, there is a Doppler shift of the generated light.  In order to have the excitation light on resonance with the $5d_{5/2}$ F=3 state, the 778 nm excitation laser must Doppler shift 14.5 MHz.  Since the Doppler shift of the blue is 1.85 times this \cite{Brekke:15}, the total frequency shift expected would result in a separation between the peaks of 113.8 MHz. 

To verify the source of separation between the peaks, we replaced the $^{87}$Rb cell with a cell containing natural isotope abundances, and examined the light generated from $^{85}$Rb.  Here we measured the separation between the peaks to be $68\pm11$ MHz, where the separation was not as precisely determined as the peaks were more difficult to resolve.  Using the same logic for Doppler shifted excitation of the $5s_{1/2}$ $F=3\rightarrow5d_{5/2}$ F=4 transition and decay through the $6p_{1/2}$ F=3 state, we would expect a separation of 47.9 MHz.  

The theoretical values for the separation between peaks is in good  agreement with our experimental measurements for $^{87}$Rb, suggesting the origin of the secondary peak is from a different hyperfine level path.  For $^{85}$Rb, the separation between peaks is smaller as expected, but the value here is not in agreement with theory and more precise data will be needed.  Additionally, the length of the $^{85}$Rb cell required a non-ideal cell placement in the ring cavity, which may have resulted in increased absorption.  Even for detuned excitation, the effects of absorption in the cell are not eliminated, and this may affect the perceived center of the blue peaks.  It has been observed \cite{Akulshin:17} that the orientation of the cell and reflected beams can play a crucial role in the four-wave mixing process as well.  The 5.23 $\mu$m beam does not leave our cell, but in the future the examination of the infrared spectrum could provide further insight into the mechanism for the alternate frequency.

\section{Saturation Process}

\begin{figure}[htbp]
\centering
\includegraphics[width=9cm]{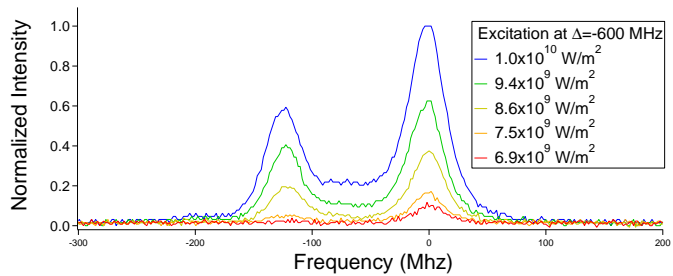}
\caption {Frequency spectrum of the 420 nm light for a variety of excitation laser intensities for excitation -600 MHz detuned from the $5s_{1/2}$ $F=2\rightarrow5d_{5/2}$ F=4 transition.  Here the ratio between the primary and secondary peak does not change as drastically as with the density dependence. }    
\label{fig:powerspectrums}
\end{figure}

The dependence of the power in the primary peak on atomic density suggests the four-wave mixing process begins to saturate.  To further consider the mechanism of saturation, the spectrum of the 420 nm light was also measured for a variety of circulating intensities for a constant density of 2.7$x10^{14}$ cm$^{-3}$, as shown in Fig. \ref{fig:powerspectrums}.  Here, a different trend between the peaks is observed, as the relative size of the secondary peak does not change as drastically.  The power produced on the primary peak continues to increase, revealing a different dependence than that observed for atomic density.  

To further understand the saturation process, we consider the interference of two alternate excitation pathways \cite{Wunderlich:90}.  As the blue and infrared fields grow, a second two-photon pathway along the $5s_{1/2}\rightarrow6p_{3/2} \rightarrow5d_{5/2}$ path interferes with the original four-wave mixing excitation.  When the two excitation rates along these paths become equal, the rate of blue production will be the same as that of blue absorption, and blue light will no longer increase as the density increases.  This saturation point is given by
\begin{equation}
\frac{\Omega_{01} \Omega_{12}}{\Delta_{1}}=\frac{\Omega_{03} \Omega_{32}}{\Delta_{3}},
\label{eqn:pathways}
\end{equation}
where $\Omega_{ab}$ is the Rabi frequency between two states, and $\Delta_{a}$ is the detuning from state a.  For the excitation scheme used here, the very large detuning from the 5p state means that the process can approach saturation even for $\mu$W level blue powers.  Since the four-wave mixing process is velocity selective, the detuning from the $5d$ state will remain zero, and does not effect the saturation process.  The detuning from the 6p state is not directly controlled in the parametric process, so the photons through the 6p state are near-resonance, and the two-photon excitation is limited by the width of the 6p state.  The number of photons for the infrared and blue are identical, which allows the saturation condition to be written in terms of the dipole moments, detunings, and wavelengths involved as       
\begin{equation}
P_{420}=\sqrt{\frac{\lambda_{32}}{\lambda_{03}}} \frac{\mu_{01}\mu_{12}}{\mu_{03}\mu_{32}} \frac{\Delta_{3}}{\Delta_{1}}P_{778}.
\label{eqn:saturation}
\end{equation}

The saturation process would then still result in an increased blue output as the input light increases.  This is further shaped by the linewidth of the $5s_{1/2}\rightarrow6p_{3/2}$ transition, as the 420 nm light produced in this process is sufficient to cause power broadening.  In the limit of 420 intensities much higher than the saturation intensity, the linewidth will go as $\sqrt{P_{420}}$.  Putting this into Eqn. \ref{eqn:saturation}, we see that the power at 420 nm is expected to scale as the square of the input power in the saturated and power broadened regime.

\begin{figure}[htbp]
\centering
\includegraphics[width=8.2cm]{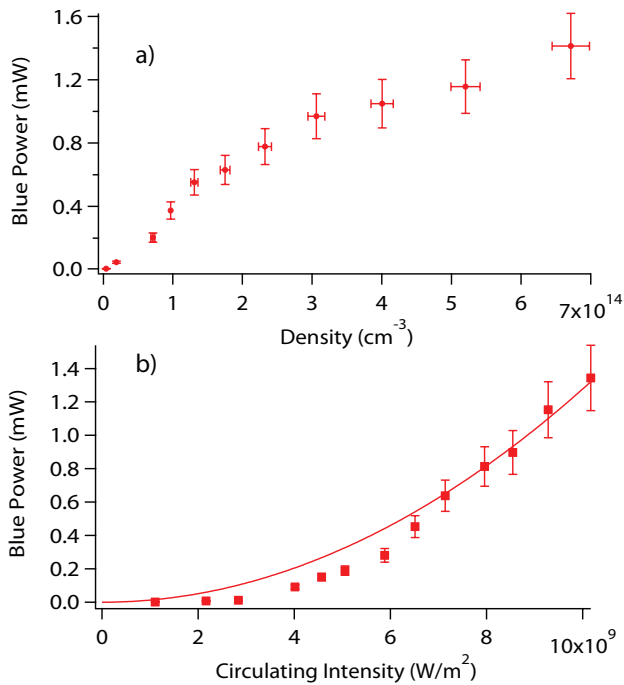}
\caption { a) The blue power is measured as a function of the atomic density in the cell, showing the output power reaches a point where gain is significantly reduced.  b) The blue power is measured as a function of the excitation laser intensity, showing a continued growth even in the high power regime. The data is fit to a quadratic dependence, showing good agreement with the expected scaling in the high power region. }
\label{fig:powerdependences}
\end{figure}

The light at the focus of the cavity inside the rubidium cell has a waist of $22 ~\mu$m, meaning that only $1 ~\mu$W of blue power would give a width of 6.5 MHz.  For the powers obtained here, linewidths in the range of 20 to 80 MHz were seen.  These linewidths are dominated by power broadening, but when the blue light is near resonance the linewidth is further shaped by the absorption of this light through the cell.  It has previously been observed that the implementation of a ring cavity resonant with the resulting blue light can both increase the generated power and dramatically reduce the linewidth below the power broadened values.  \cite{Offer:16}

The dependence of the generated 420 nm light on both atomic density and excitation power is shown in Fig. \ref{fig:powerdependences}.  This can now be understood in terms of the saturation process. As the blue power generated on the F=4 hyperfine pathway increases, the rate of blue generation via FWM can become equal to the rate of blue light used for two photon excitation through the 6p state.  This would result in no further increase in generated blue light along the primary pathway with increasing density.  At the same time, the generation of blue light on the alternate F=3 hyperfine pathway would continue.  In Fig. \ref{fig:powerdependences}a the growth in blue power is significantly reduced at high densities, suggesting only a limited gain in power due to the secondary pathway.

Fig. \ref{fig:powerdependences}b shows that blue power continues to grow with input intensity. Here the trend is expected to be  quadratic growth once the process has reached saturation.  The data is fit to a quadratic dependence, showing good agreement in the high power region.  This further suggests the regime where the excitation rate along the 6p path has become equal to the original excitation rate along the 5p path.

The presence of this saturation process on the primary four-wave mixing process can also provide insight on the two frequency components observed.  Four-wave mixing through the $5d_{5/2}$ F=4 state is much more efficient than through the $5d_{5/2}$ F=3 state, and so dominates the generated power for low densities and input powers.  However, as density increases the process through the $5d_{5/2}$ F=4 state approaches saturation while the process through the $5d_{5/2}$ F=3 state continues to grow.  As seen here, this can then result in these two processes generating comparable powers.

\section{Discussion and Outlook}

We have investigated the frequency spectrum and saturation characteristics of 420 nm light generated through parametric four-wave mixing in rubidium vapor.  Power produced along the primary pathway saturates when the rate for excitation along the $5s_{1/2}\rightarrow6p_{3/2} \rightarrow5d_{5/2}$ becomes comparable to the $5s_{1/2}\rightarrow5p_{3/2} \rightarrow5d_{5/2}$ pathway.  For the two-photon excitation far detuned from the 5p state, the rates along these two paths can become equal for $\mu$Ws of blue light.  At this point higher densities do not result in further blue light production, though a quadratic dependence on excitation intensity remains.  

Under low power conditions, the four-wave mixing process results in a single generated frequency, but as saturation approaches a second peak becomes increasingly significant.  It is seen that though originally dominated by excitation to the $5d_{5/2}$ F=4 state, an alternate four wave mixing path through the $5d_{5/2}$ F=3$\rightarrow5p_{3/2}$ F=2 becomes increasingly important as the primary transition saturates.  

The onset of saturation will be delayed by smaller detunings from the 5p state, such as in the near resonant two-step process.  It could also be delayed by controlling the detuning of the resulting light from the 6p state, perhaps by seeding the four-wave mixing process \cite{Gai:16}.  Care in the selection of excitation parameters is necessary to minimize the importance of saturation and limit the effects of alternate channels to ensure the generation of single frequency beams in these systems.

\bibliography{saturation}

\end{document}